\documentclass[12pt]{article}
\usepackage{amssymb,epsfig}
\def\d{\partial}
\def\l{\left(}
\def\r{\right)}

\newcommand{\be}{\begin{equation}}
\newcommand{\ee}{\end{equation}}

\newcommand{\bg}{\begin{gather}}
\newcommand{\eg}{\end{gather}}
\def\half{\frac{1}{2}}

\begin{document}

\begin{center}
{\Large\bf Prospects for sgoldstino search at the LHC}\\
\vspace{0.3cm}
D.~S.~Gorbunov\footnote{{\bf e-mail}: gorby@ms2.inr.ac.ru}, 
N.~V.~Krasnikov\footnote{{\bf e-mail}: krasniko@ms2.inr.ac.ru}, \\
{\small{\em
Institute for Nuclear Research of the Russian Academy of Sciences, }}\\
{\small{\em
60th October Anniversary prospect 7a, Moscow 117312, Russia
}}
\end{center}
\begin{abstract}  
In this paper we estimate the LHC  sgoldstino discovery 
potential for the signatures with $ \gamma \gamma$ and 
$Z Z$ in a final state. 

\end{abstract} 


It is well known, that exist models of supergravity breaking with 
relatively light sgoldstinos (scalar $S$ and pseudoscalar $P$ particles
---  superpartners of goldstino $\psi$). 
Such pattern emerges in a number of non-minimal 
supergravity models~\cite{ellis} and also in gauge mediation
models if supersymmetry is broken via non-trivial superpotential (see,
Ref.~\cite{gmm} and references therein).
To the leading order in $1/F$, where
$F$ is the parameter of supersymmetry breaking, and to 
zero order in MSSM gauge
and Yukawa coupling constants, the interactions between 
the component fields of
goldstino supermultiplet and MSSM fields have been derived in
 Ref.~\cite{Gorbunov:2001pd}. 
They correspond to the most attractive for collider studies 
processes where only one of these {\it new} particles appears in a
 final state. In this case light gravitino behaves exactly
as goldstino. For sgoldstinos, as they are R-even, 
only sgoldstino couplings to goldstino and 
sgoldstino couplings to SM fields have been included as the
 most interesting phenomenologically. 

All relevant sgoldstino coupling constants 
presented in Ref.~\cite{Gorbunov:2001pd} are
completely determined by the MSSM soft terms and the parameter of
supersymmetry breaking $F$, but sgoldstino masses ($m_S,m_P$) remain
free. If sgoldstino masses are of the order of electroweak scale and 
$\sqrt{F}\sim1$~TeV --- sgoldstino may be detected 
in collisions of high energy particles at 
supercolliders~\cite{Perazzi:2000id,Perazzi:2000ty}.    

There are flavor-conserving and flavor-violating interactions of
sgoldstino fields.
As concerns flavor-conserving interactions, the strongest bounds arise
from astrophysics and cosmology, that is $\sqrt{F}\gtrsim
10^6$~GeV~\cite{Nowakowski:1994ag,Gorbunov:2000th}, or
$m_{3/2}>600$~eV, for models with $m_{S(P)}<10$~keV and MSSM soft
flavor-conserving terms being of the order of electroweak scale.  For
the intermediate sgoldstino masses (up to a few MeV) constraints from
the study of SN explosions and reactor experiments lead to
$\sqrt{F}\gtrsim300$~TeV~\cite{Gorbunov:2000th}.  For heavier
sgoldstinos, low energy processes (such as rare decays of mesons)
provide limits at the level of
$\sqrt{F}\gtrsim500$~GeV~\cite{Gorbunov:2000th}.

The collider experiments exhibit the same level of sensitivity to
light sgoldstinos as rare meson decays.  Indeed the
studies~\cite{Dicus:1989gg,Dicus:1990su,Dicus:1990dy,Dicus:1996ua} of
the light sgoldstino ($m_{S,P}\lesssim a~few$~MeV) phenomenology based
on the effective low-energy Lagrangian derived from N=1 linear
supergravity yield the bounds: $\sqrt{F}\gtrsim500$~GeV (combined
bound on $Z\to S\bar{f}f,P\bar{f}f$~\cite{Dicus:1990dy}; combined
bound on $e^+e^-\to\gamma S,\gamma P$~\cite{Dicus:1990su}) at
$M_{soft}\sim100$~GeV, $\sqrt{F}\gtrsim1$~TeV~\cite{Dicus:1996ua}
(combined bound on $p\bar{p}\to gS,gP$) at gluino mass $M_3\simeq500$~GeV.
Searches for heavier sgoldstinos at colliders, though exploiting
another technique, results in similar bounds on the scale of
supersymmetry breaking.  Most powerful among the operating machines,
LEP and Tevatron, give a constraint of the order of 1~TeV on
supersymmetry breaking scale in models with light sgoldstinos. Indeed,
the analysis carried out by DELPHI Collaboration~\cite{Abreu:2000ij}
yields the limit $\sqrt{F}>500\div200$~GeV at sgoldstino masses
$m_{S,P}=10\div150$~GeV and $M_{soft}\sim100$~GeV.  The constraint
depends on the MSSM soft breaking parameters. In particular, it is
stronger by about a hundred GeV in the model with degenerate gauginos.
At Tevatron, a few events in $p\bar{p}\to S\gamma(Z)$ channel, and
about $10^4$ events in $p\bar{p}\to S$ channel would be produced at
$\sqrt{F}=1$~TeV and $M_{soft}\sim100$~GeV for integrated luminosity
${\cal L}=100$~pb$^{-1}$ and sgoldstino mass of the order of
100~GeV~\cite{Perazzi:2000ty}.  This gives rise to a possibility to
detect sgoldstino, if it decays inside the detector into photons and
$\sqrt{F}$ is not larger than $1.5\div2$~TeV.

In this note we estimate  the LHC sgoldstino discovery potential 
using as  a signature the decay of sgoldstino into two photons 
or(and) two Z-bosons. 

In terms of $SU(3)_c\times SU(2)_L\times U(1)_Y$ fields the sgoldstino
effective lagrangian reads~\cite{Gorbunov:2001pd}:
\begin{eqnarray}
{\cal L}_{S}&=&-\sum_{all~gauge\atop fields}
{M_\alpha\over2\sqrt{2}F}S\cdot F_{a~\mu\nu}^\alpha F_a^{\alpha~\mu\nu}
-{{\cal A}^L_{ab}\over\sqrt{2} F}y^L_{ab}\cdot S
\bigl(\epsilon_{ij} l_a^je_b^c h_D^i +h.c.\bigr)
\nonumber
\\&-&
{{\cal A}_{ab}^D\over\sqrt{2} F}y_{ab}^D\cdot S
\bigl( \epsilon_{ij} q_a^jd_b^c h_D^i+h.c.\bigr)
-{{\cal A}_{ab}^U\over\sqrt{2} F}y_{ab}^U\cdot S
\bigl( \epsilon_{ij} q_a^iu_b^ch_U^j+h.c.\bigr)\;,
\nonumber
\\
{\cal L}_{P}&=&\sum_{all~gauge\atop fields}
{M_\alpha\over 4\sqrt{2}F}P\cdot F_{a~\mu\nu}^\alpha 
\epsilon^{\mu\nu\lambda\rho}F_{a~\lambda\rho}^\alpha
-i{{\cal A}^L_{ab}\over\sqrt{2} F}y^L_{ab}\cdot P
\bigl(\epsilon_{ij} l_a^je_b^c h_D^i -h.c.\bigr)
\nonumber
\\&-&
i{{\cal A}_{ab}^D\over\sqrt{2} F}y_{ab}^D\cdot P
\bigl( \epsilon_{ij} q_a^jd_b^c h_D^i-h.c.\bigr)
-i{{\cal A}_{ab}^U\over\sqrt{2} F}y_{ab}^U\cdot P
\bigl( \epsilon_{ij} q_a^iu_b^ch_U^j-h.c.\bigr)\;.
\nonumber
\\
{\cal L}_{\psi,S,P}&=&i\d_\mu\bar{\psi}\bar{\sigma}^\mu\psi
+\half\d_\mu S\d^\mu S
-\half m_S^2S^2+\half\d_\mu P\d^\mu P-\half m_P^2P^2
\nonumber
\\
&+&{m_S^2\over2\sqrt{2}F}S\l\psi\psi+\bar{\psi}\bar{\psi}\r
-i{m_P^2\over2\sqrt{2}F}P\l\psi\psi-\bar{\psi}\bar{\psi}\r\;.
\nonumber
\end{eqnarray}  
where $M_\alpha$ are
gaugino masses and $A_{\alpha\beta}y_{\alpha\beta}$ are soft trilinear
coupling constants. In this letter we consider ${\cal A}_{ab}=A$ and Yukawas
$y_{ab}\propto\delta_{ab}$ as in SM. 

At hadron colliders sgoldstinos will be produced mostly by gluon
resonant scattering $gg\to S(P)$~\cite{Perazzi:2000ty}. The associated
production $gg\to S(P)g$ has several times smaller cross section than
resonant production and the corresponding discovery potential (in
analogy with SM Higgs boson case) is expected to be weaker than the
discovery potential for the resonant mode $gg\to S(P)$. 

One has to consider the
subsequent decay of the sgoldstino inside the detector.
Indeed, for the range of parameters that are relevant for this study,
sgoldstinos are expected to decay inside the detector, not
far from the collision point. Then, assuming that the
supersymmetric partners (others than the gravitino $\tilde G$) are too
heavy to be relevant for the sgoldstino decays, the main decay channels are:
\[
S(P)\to gg, \gamma\gamma, \tilde{G}\tilde{G}, f \bar f, \gamma Z, WW, ZZ.
\]
The corresponding widths have been calculated 
in Refs.~\cite{Perazzi:2000id,Perazzi:2000ty}~\footnote{There is an
additional parameter $\mu_a$ in sgoldstino decay widths into weak bosons 
presented in Ref.~\cite{Perazzi:2000id}; this parameter is absent in
the minimal model considered in this letter, see 
Ref.~\cite{Gorbunov:2001pd}.}.
                                    
For a  sgoldstinos decaying into pairs of massless gauge bosons, one has
\[
\Gamma(S(P)\to\gamma\gamma)={M_{\gamma\gamma}^2m_{S(P)}^3\over 32\pi F^2}\;,
~~~~
\Gamma(S(P)\to gg)={M_3^2m_{S(P)}^3\over 4\pi F^2}\;,
\]
where $M_{\gamma\gamma}=M_1\cos^2\theta_W+M_2\sin^2\theta_W$,
and $\theta_W$ is the electroweak mixing angle. Note that for
$M_{\gamma\gamma}\sim M_3$ gluonic mode dominates over the photonic 
one due to the color factor enhancement.
  
For the values of $\sqrt{F}$ we are interested in, gravitino is very light,
with mass in the range $m_{\tilde{G}}=\sqrt{8\pi/3}\;
F/M_{Pl}\simeq10^{-3}\div 10^{-1}$~eV. Then, the sgoldstino decay rates
into two gravitinos are given by
\[
\Gamma(S(P)\to\tilde{G}\tilde{G})={m_{S(P)}^5\over 32\pi F^2}\;,
\]
and become comparable with the rate into two photons for heavy
sgoldstinos, such that $m_{S(P)}\sim M_{\gamma\gamma}$. 
Sgoldstinos can also decay into fermion pairs, with rates
\begin{eqnarray}
\Gamma(S\to f\bar{f})=N_C{A^2m_f^2m_S\over 32\pi
F^2}\l1-{4m_f^2\over m_S^2}\r^{3/2}\;,
\nonumber\\
\Gamma(P\to f\bar{f})=N_C{A^2m_f^2m_P\over 32\pi
F^2}\l1-{4m_f^2\over m_P^2}\r^{1/2}\;,
\nonumber
\end{eqnarray}
where $m_f$ is fermion mass, and $N_C=3$ for quarks
and $N_C=1$ for leptons. One can see that, far from the threshold, the
fermionic branching ratios are suppressed by a factor $m_f^2/m_S^2$ in general. Hence,
the decay $S(P) \to f \bar f$ can be relevant  for large trilinear
couplings and/or if the sgoldstino mass happens to be not too far from
$m_f$. Finally, sgoldstinos lighter than the top quark can decay
into massive vector bosons states. For $m_{S(P)} > M_Z$,  
$m_{S(P)} > 2M_W$ and $m_{S(P)} > 2M_Z$ 
the $Z\gamma$, $W^+W^-$ and $ZZ$ channels open up, respectively. 
The corresponding rates read
\begin{eqnarray}
\Gamma(S(P)\to \gamma Z)&=&{M_{\gamma Z}^2m_{S(P)}^3\over 16\pi
F^2}\l1-{M_Z^2\over m_{S(P)}^2}\r^3\!\!,
\nonumber\\
\Gamma(P\to W^+W^-)&=&{M_2^2m_{P}^3\over 16\pi F^2}
\l1-{4M_W^2\over m_{P}^2}\r^{3/2}\!\!\!\!\!\!,
\nonumber\\
\Gamma(S\to W^+W^-)&=&{M_2^2m_S^3 \over 16\pi F^2}\l
1-4{M_W^2\over m_S^2}+6{M_W^4\over m_S^4}\r
\sqrt{1-{4M_W^2\over m_{S}^2}}\;,
\nonumber\\
\Gamma(P\to ZZ)&=&{M_{ZZ}^2m_{P}^3\over 32\pi F^2}
\l1-{4M_Z^2\over m_{P}^2}\r^{3/2}\!\!\!\!\!\!,
\nonumber\\
\Gamma(S\to ZZ)&=&{M_{ZZ}^2m_S^3 \over 32\pi F^2}\l
1-4{M_Z^2\over m_S^2}+6{M_Z^4\over m_S^4}\r
\sqrt{1-{4M_Z^2\over m_{S}^2}}\;,
\nonumber
\end{eqnarray} 
where $M_{\gamma Z}=(M_2-M_1)\cos\theta_W\sin\theta_W$ and 
$M_{ZZ}=M_1\sin^2\theta_W+M_2\cos^2\theta_W$.

\begin{figure}
\centerline{\epsfxsize=1.0\textwidth \epsfbox{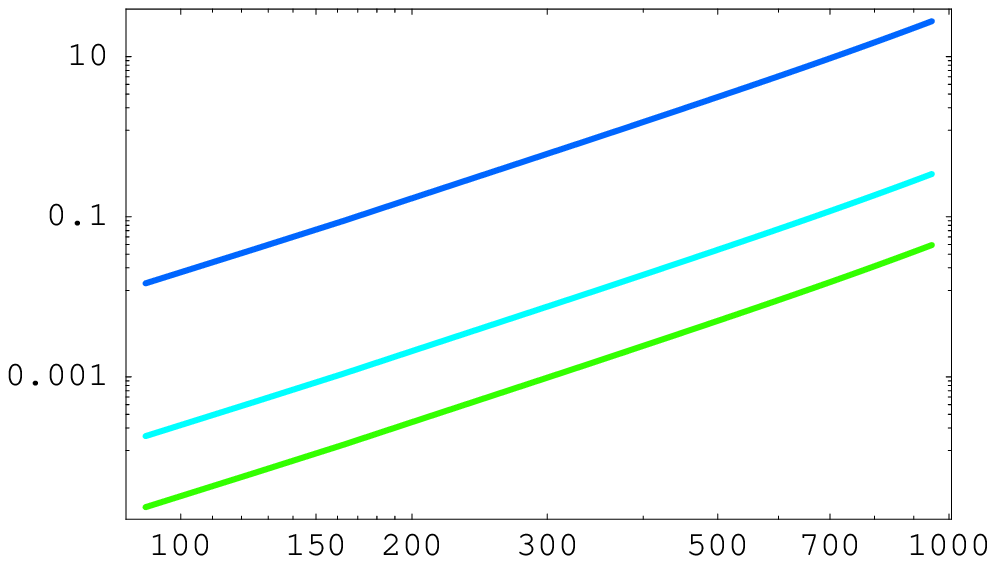}}
\caption{Total decay width of sgoldstino  
$\Gamma_S$ as a function of its mass $m_S$ 
for the model~I.
\label{fig:width-1}}
\begin{picture}(0,0)(-28,76)
\put(-30.00,370.00){\makebox(0,0)[cb]{\Large $\Gamma_S$,~GeV}}
\put(260.00,352.00){\makebox(0,0)[cb]{$\sqrt{F}=1$~TeV}}
\put(260.00,285.00){\makebox(0,0)[cb]{$\sqrt{F}=3$~TeV}}
\put(260.00,215.00){\makebox(0,0)[cb]{$\sqrt{F}=5$~TeV}}
\put(200.00,125.00){\makebox(0,0)[cb]{\large$m_S$,~GeV}}
\end{picture}
\end{figure}

\begin{figure}[htb]
\centerline{\epsfxsize=1.0\textwidth \epsfbox{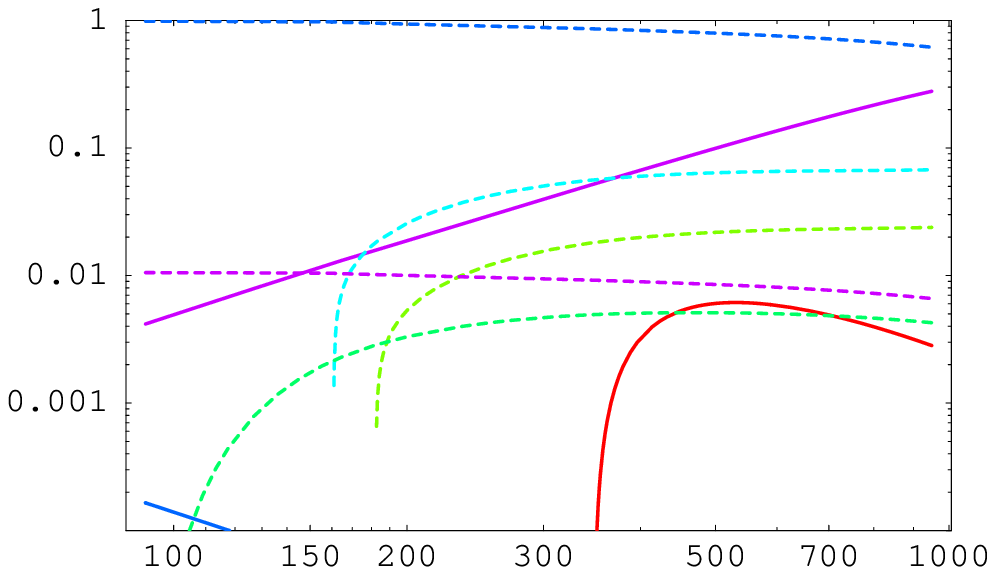}}
\caption{Sgoldstino branching ratios Br$_S$ as functions of its mass
$m_S$ for the model~I. 
\label{branchings-1}}
\begin{picture}(0,0)(-37,-244)
\put(-35.00,30.00){\makebox(0,0)[cb]{\Large Br$_S$}}
\put(295.00,37.00){\makebox(0,0)[cb]{$gg$}}
\put(33.00,-147.00){\makebox(0,0)[cb]{$b\bar{b}$}}
\put(75.00,-105.00){\makebox(0,0)[cb]{$\gamma Z$}}
\put(40.00,-85.00){\makebox(0,0)[cb]{$\tilde{G}\tilde{G}$}}
\put(40.00,-45.00){\makebox(0,0)[cb]{$\gamma\gamma$}}
\put(235.00,-120.00){\makebox(0,0)[cb]{$t\bar{t}$}}
\put(132.00,-25.00){\makebox(0,0)[cb]{$W^+W^-$}}
\put(125.00,-133.00){\makebox(0,0)[cb]{$ZZ$}}
\put(200.00,-197.00){\makebox(0,0)[cb]{\large$m_S$,~GeV}}
\end{picture}
\end{figure}

\begin{figure}[htb]
\centerline{\epsfxsize=1.0\textwidth \epsfbox{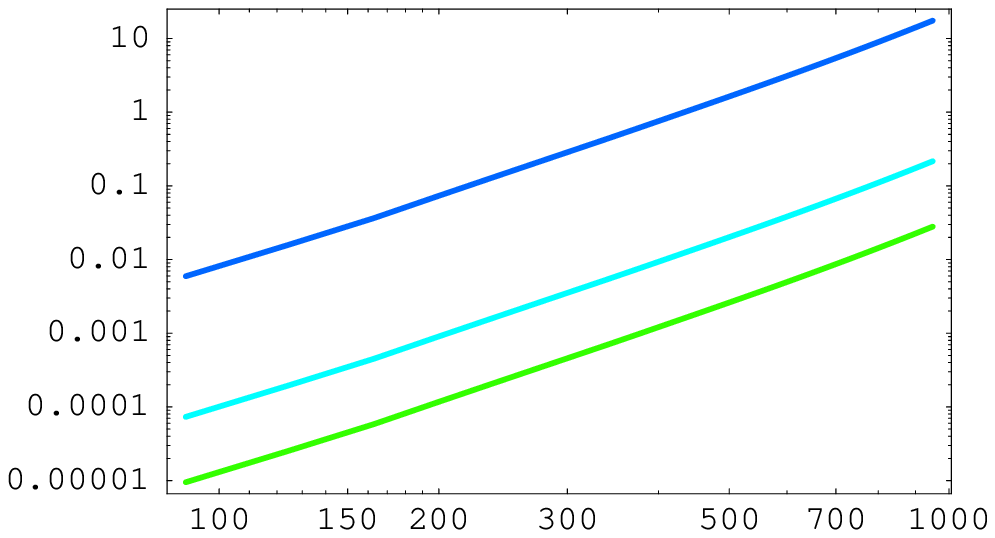}}%
\caption{Total decay width of sgoldstino
$\Gamma_S$ as a function of its mass $m_S$ for the model~II.
\label{fig:width-2}}
\begin{picture}(0,0)(-38,71)%
\put(-30.00,370.00){\makebox(0,0)[cb]{\Large $\Gamma_S$,~GeV}}
\put(260.00,344.00){\makebox(0,0)[cb]{$\sqrt{F}=1$~TeV}}
\put(260.00,282.00){\makebox(0,0)[cb]{$\sqrt{F}=3$~TeV}}
\put(260.00,215.00){\makebox(0,0)[cb]{$\sqrt{F}=5$~TeV}}
\put(200.00,125.00){\makebox(0,0)[cb]{\large$m_S$,~GeV}}
\end{picture}%
\end{figure}

\begin{figure}[htb]
\centerline{\epsfxsize=1.0\textwidth \epsfbox{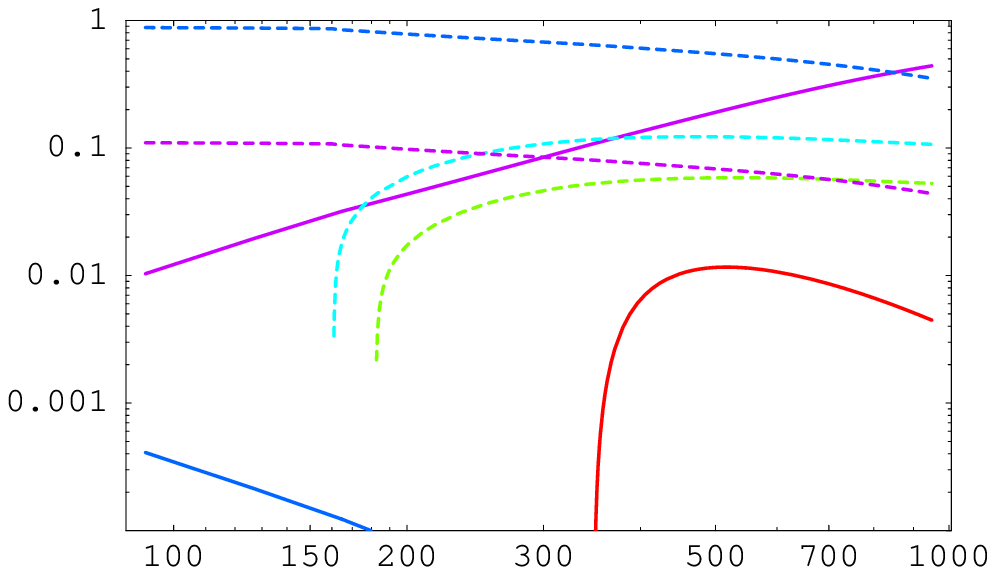}}
\caption{Sgoldstino branching ratios Br$_S$ as functions of its mass
$m_S$ for the model~II.
\label{branchings-2}}
\begin{picture}(0,0)(-38,-246)%
\put(-35.00,30.00){\makebox(0,0)[cb]{\Large Br$_S$}}
\put(40.00,42.00){\makebox(0,0)[cb]{$gg$}}
\put(35.00,-125.00){\makebox(0,0)[cb]{$b\bar{b}$}}
\put(40.00,-65.00){\makebox(0,0)[cb]{$\tilde{G}\tilde{G}$}}
\put(40.00,-10.00){\makebox(0,0)[cb]{$\gamma\gamma$}}
\put(235.00,-120.00){\makebox(0,0)[cb]{$t\bar{t}$}}
\put(87.00,-70.00){\makebox(0,0)[cb]{$W^+W^-$}}
\put(127.00,-105.00){\makebox(0,0)[cb]{$ZZ$}}
\put(200.00,-197.00){\makebox(0,0)[cb]{\large$m_S$,~GeV}}
\end{picture}%
\end{figure}

We will present the estimates for the LHC sensitivity to the scale of 
supersymmetry breaking for two sets of MSSM soft parameters shown 
in Table~\ref{sets}. 
\begin{table}[htb]
\begin{center}
\vspace{2mm}  
\begin{tabular}{|c|c|c|c|c|}     
\hline
Model&$M_1$&$M_2$&$M_3$&$A$\\  
\hline
I&100~GeV&300~GeV&500~GeV&300~GeV\\
\hline
II&300~GeV&300~GeV&300~GeV&300~GeV\\
\hline
\end{tabular}
\caption{The sets of parameters which the LHC sensitivity is presented
for.
\label{sets}}
\end{center}
\end{table} 
For these sets of parameters we calculate sgoldstino width 
(see Figures~\ref{fig:width-1},\ref{fig:width-2}) and
branching ratios (see Figures~\ref{branchings-1},\ref{branchings-2}). 
In fact, only $gg$, $\gamma\gamma$, $ZZ$, $W^+W^-$ and
$\tilde{G}\tilde{G}$ modes are relevant in our study.


For  $\gamma\gamma$ mode the simulations of the CMS
detector~\cite{CMS-TP} lead for the Higgs boson masses 
$m_{h} = 100, 110, 130$~GeV to the mass resolutions  $\Delta M = 0.78,
0.87, 0.96$~GeV
(high luminosity $L = 10^{34}$~cm$^{-2}$s$^{-1}$). 
So, to estimate the 
diphoton mass resolution we shall use the simplest parametrisation
\[
{\Delta M\over M} = 0.008\;. 
\] 
In order to estimate the 4 lepton mass resolution 
$ZZ \rightarrow 4~leptons$ 
we use the parametrisation
\[
{\Delta M\over M} = 0.02\;. 
\] 

Defining a signal significance as 
\[
\Sigma={N_{\rm S}\over \sqrt{N_{\rm B}}}=
{\sigma_{\rm S}\over\sqrt{\sigma_{\rm B}}}\sqrt{{\cal L}^{-1}_{\rm LHC}}\;,
\]
where ${\cal L}^{-1}_{\rm LHC}$ is the integrated luminosity of LHC, we
estimate the signal significance for sgoldstino events: 
\begin{equation}
\Sigma_S=\Sigma_h\cdot{\Gamma(S\to gg)\over\Gamma(h\to gg)}
~{{\rm Br}_S^{\gamma\gamma(ZZ)}\over{\rm Br}_h^{\gamma\gamma(ZZ)}}
~\sqrt{{\rm max}(\Gamma_S,\Delta M) \over {\rm max}(\Gamma_h,\Delta M)}\;,
\label{estimate-of-significance}
\end{equation}
with $\Sigma_h$ being the signal significance for SM Higgs boson. We
will exploit only $\gamma\gamma$ and $ZZ$ channels. In
what follows we will use $\Sigma_h(m_h)$ presented by CMS
Collaboration in
Refs.~\cite{signif-higgs-gamma-gamma,Denegri:1995fr} 
for $\gamma\gamma$ and $ZZ$ channels, respectively. 
Note that partial width for both scalars (sgoldstino and Higgs boson) 
should be calculated either in the leading order or with account of 
electroweak/strong corrections. The results coincide, since the
relevant corrections for any neutral scalar SM singlet are the same. 

For two sets of the MSSM soft breaking terms shown in Table~\ref{sets} 
we present plots with lines of the LHC sensitivities to 
sgoldstinos of various masses. 
 
\begin{figure}[htb]
\centerline{\epsfxsize=1.0\textwidth \epsfbox{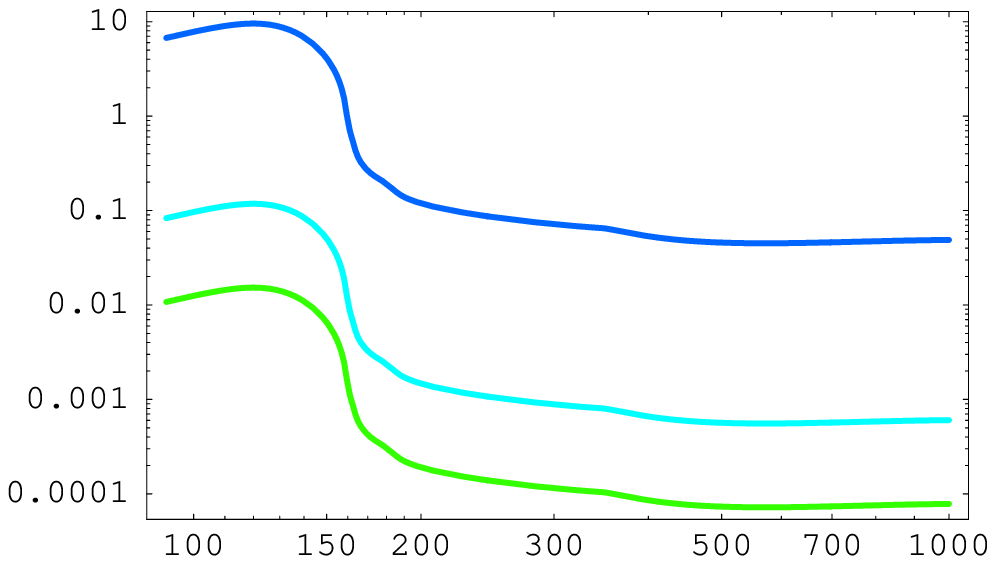}}
\caption{The ratio of sgoldstino width $\Gamma_S$ and 
Higgs width $\Gamma_h$ as a function of mass $m$ at various
$\sqrt{F}$ for the model~I.
\label{ratio-1}}
\begin{picture}(0,0)(-38,106)
\put(-30.00,350.00){\makebox(0,0)[cb]{\Large $\Gamma_S/\Gamma_h$}}
\put(320.00,317.00){\makebox(0,0)[cb]{\large$\sqrt{F}=1$~TeV}}
\put(320.00,238.00){\makebox(0,0)[cb]{\large$\sqrt{F}=3$~TeV}}
\put(320.00,202.00){\makebox(0,0)[cb]{\large$\sqrt{F}=5$~TeV}}
\put(200.00,155.00){\makebox(0,0)[cb]{\large$m$,~GeV}}
\end{picture}
\end{figure}

\begin{figure}[htb]
\centerline{\epsfxsize=1.0\textwidth \epsfbox{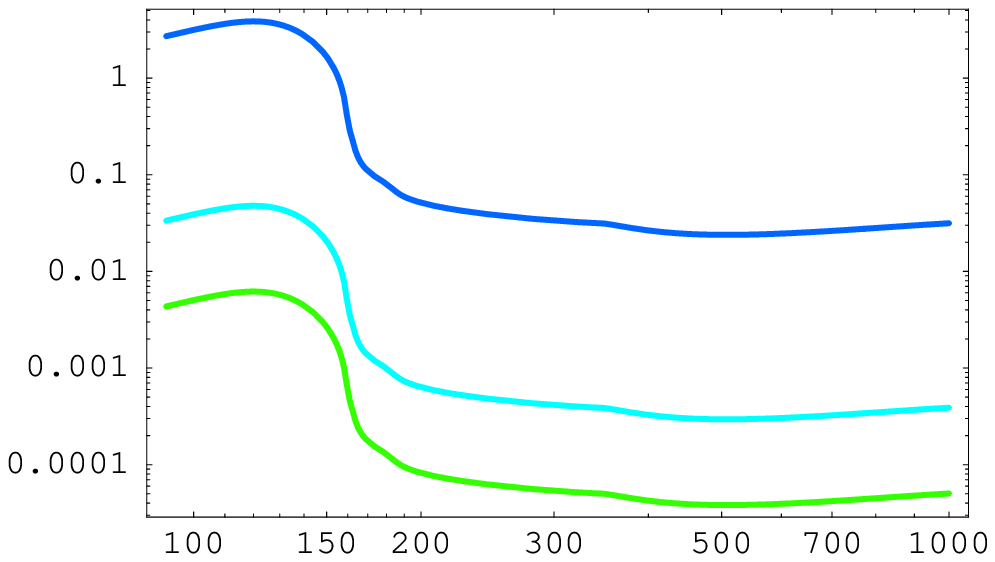}}
\nonumber
\caption{The ratio of sgoldstino width $\Gamma_S$ and Higgs width $\Gamma_h$ as
a function of mass $m$ at various $\sqrt{F}$ for the model~II.
\label{ratio-2}}
\begin{picture}(0,0)(-38,106)
\put(-30.00,362.00){\makebox(0,0)[cb]{\Large $\Gamma_S/\Gamma_h$}}
\put(320.00,323.00){\makebox(0,0)[cb]{\large$\sqrt{F}=1$~TeV}}
\put(320.00,243.00){\makebox(0,0)[cb]{\large$\sqrt{F}=3$~TeV}}
\put(320.00,205.00){\makebox(0,0)[cb]{\large$\sqrt{F}=5$~TeV}}
\put(200.00,155.00){\makebox(0,0)[cb]{\large$m$,~GeV}}
\end{picture}
\end{figure}

First, we present the plots~\ref{ratio-1},\ref{ratio-2} 
with the ratios of sgoldstino-to-Higgs widths. One can see, that for
the relevant region in ($m_{S(P)},\sqrt{F}$) space, sgoldstino width is 
more narrow. 

\begin{figure}[htb]
\centerline{\epsfxsize=1.0\textwidth \epsfbox{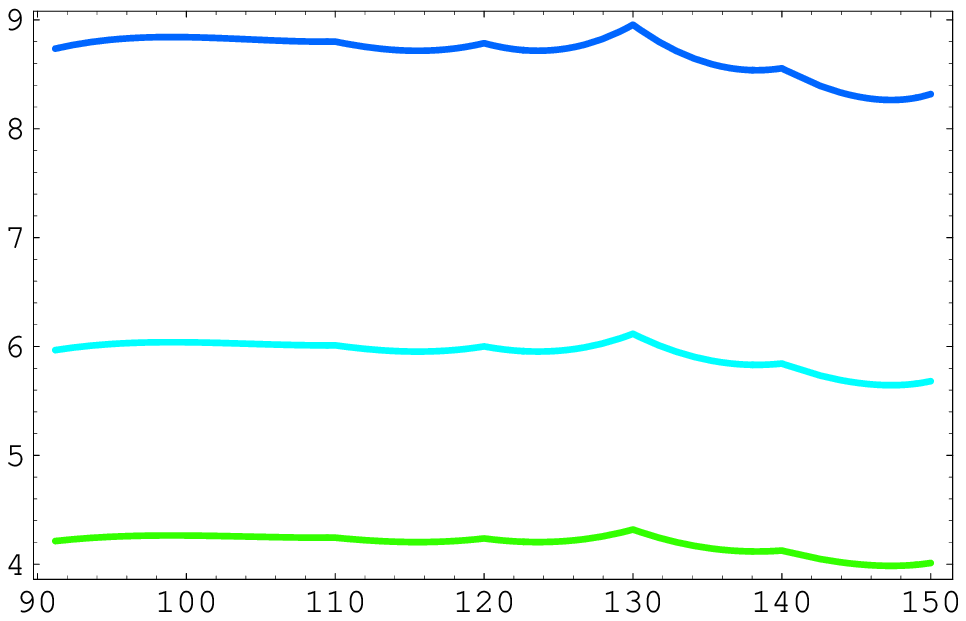}}
\caption{Signal significance of $\gamma\gamma$ channel 
as a function of sgoldstino mass $m_S$ for the model~I.
\label{fig:photon-1}}
\begin{picture}(0,0)(-38,224)
\put(-50.00,505.00){\makebox(0,0)[cb]{\LARGE$\frac{N_{\rm
S}}{\sqrt{N_{\rm B}}}$}}
\put(40.00,465.00){\makebox(0,0)[cb]{${\cal L}=100$~fb$^{-1}$}}
\put(320.00,485.00){\makebox(0,0)[cb]{\large$\sqrt{F}=5$~TeV}}
\put(320.00,395.00){\makebox(0,0)[cb]{\large$\sqrt{F}=5.5$~TeV}}
\put(320.00,315.00){\makebox(0,0)[cb]{\large$\sqrt{F}=6$~TeV}}
\put(200.00,265.00){\makebox(0,0)[cb]{\large$m_S$,~GeV}}
\end{picture}
\end{figure}

\begin{figure}[htb]
\centerline{\epsfxsize=1.0\textwidth \epsfbox{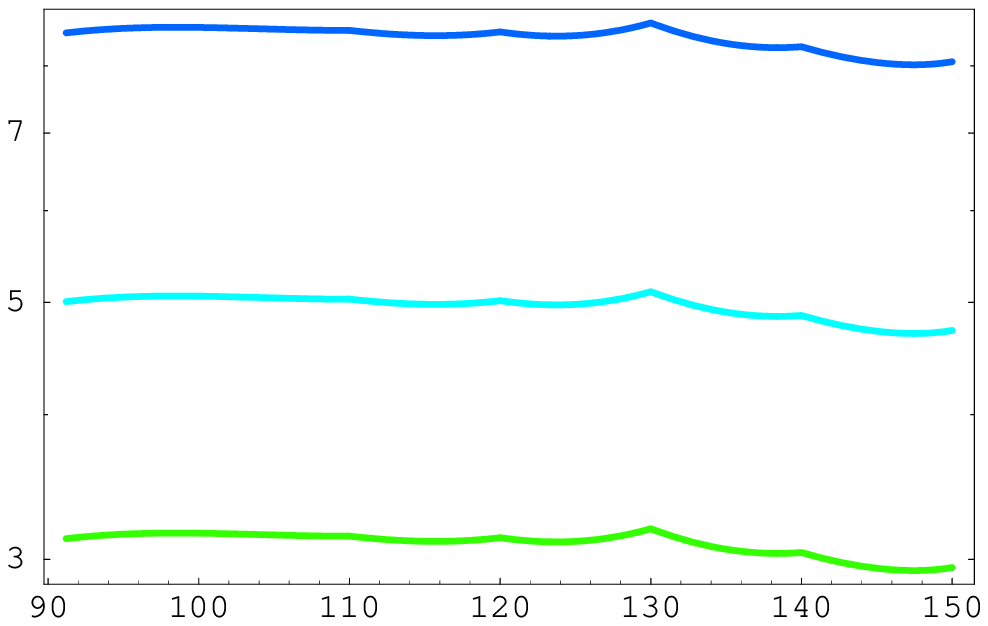}}
\caption{Signal significance of $\gamma\gamma$ channel as a function
of sgoldstino mass $m_S$ for the model~II.
\label{fig:photon-2}}
\begin{picture}(0,0)(-38,224)
\put(-50.00,512.00){\makebox(0,0)[cb]{\LARGE$\frac{N_{\rm
S}}{\sqrt{N_{\rm B}}}$}}
\put(40.00,485.00){\makebox(0,0)[cb]{${\cal L}=100$~fb$^{-1}$}}
\put(200.00,512.00){\makebox(0,0)[cb]{\large$\sqrt{F}=7$~TeV}}
\put(200.00,428.00){\makebox(0,0)[cb]{\large$\sqrt{F}=8$~TeV}}
\put(200.00,322.00){\makebox(0,0)[cb]{\large$\sqrt{F}=9$~TeV}}
\put(200.00,265.00){\makebox(0,0)[cb]{\large$m_S$,~GeV}}
\end{picture}
\end{figure}

\begin{figure}[htb]
\centerline{\epsfxsize=1.0\textwidth \epsfbox{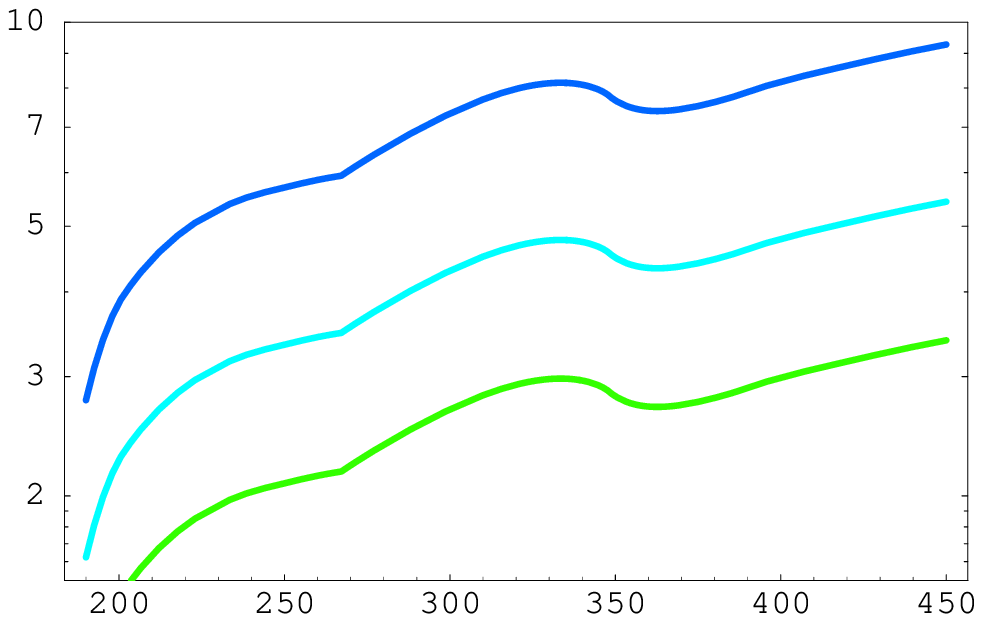}}
\caption{Signal significance of $ZZ$ channel as a function of 
sgoldstino mass $m_S$ for the model~I.
\label{fig:z-low-1}}
\begin{picture}(0,0)(-38,72)
\put(-60.00,350.00){\makebox(0,0)[cb]{\LARGE$\frac{N_{\rm
S}}{\sqrt{N_{\rm B}}}$}}
\put(40.00,355.00){\makebox(0,0)[cb]{${\cal L}=100$~fb$^{-1}$}}
\put(320.00,335.00){\makebox(0,0)[cb]{\large$\sqrt{F}=1.75$~TeV}}
\put(320.00,270.00){\makebox(0,0)[cb]{\large$\sqrt{F}=2$~TeV}}
\put(320.00,205.00){\makebox(0,0)[cb]{\large$\sqrt{F}=2.25$~TeV}}
\put(200.00,115.00){\makebox(0,0)[cb]{\large$m_S$,~GeV}}
\end{picture}
\end{figure}

\begin{figure}[htb]
\centerline{\epsfxsize=1.0\textwidth \epsfbox{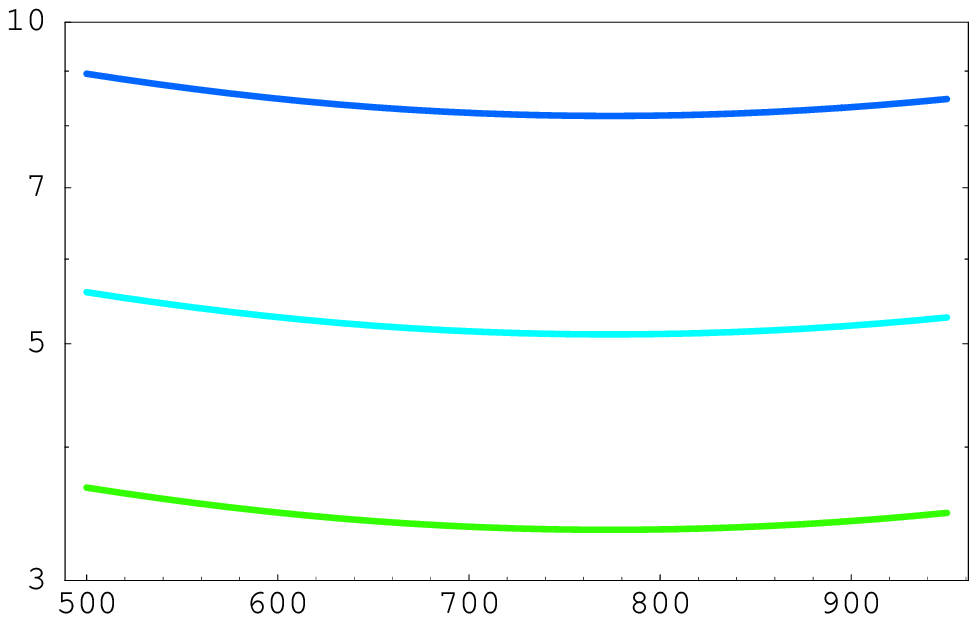}}
\caption{Signal significance of $ZZ$ channel as a function of 
sgoldstino mass $m_S$ for the model~I.
\label{fig:z-high-1}}
\begin{picture}(0,0)(-38,-248)
\put(50.00,-5.00){\makebox(0,0)[cb]{${\cal L}=100$~fb$^{-1}$}}
\put(-50.00,33.00){\makebox(0,0)[cb]{\LARGE$\frac{N_{\rm
S}}{\sqrt{N_{\rm B}}}$}}
\put(320.00,35.00){\makebox(0,0)[cb]{\large$\sqrt{F}=2$~TeV}}
\put(320.00,-59.00){\makebox(0,0)[cb]{\large$\sqrt{F}=2.25$~TeV}}
\put(320.00,-145.00){\makebox(0,0)[cb]{\large$\sqrt{F}=2.5$~TeV}}
\put(200.00,-202.00){\makebox(0,0)[cb]{\large$m_S$,~GeV}}
\end{picture}
\end{figure}

\begin{figure}[htb]
\centerline{\epsfxsize=1.0\textwidth \epsfbox{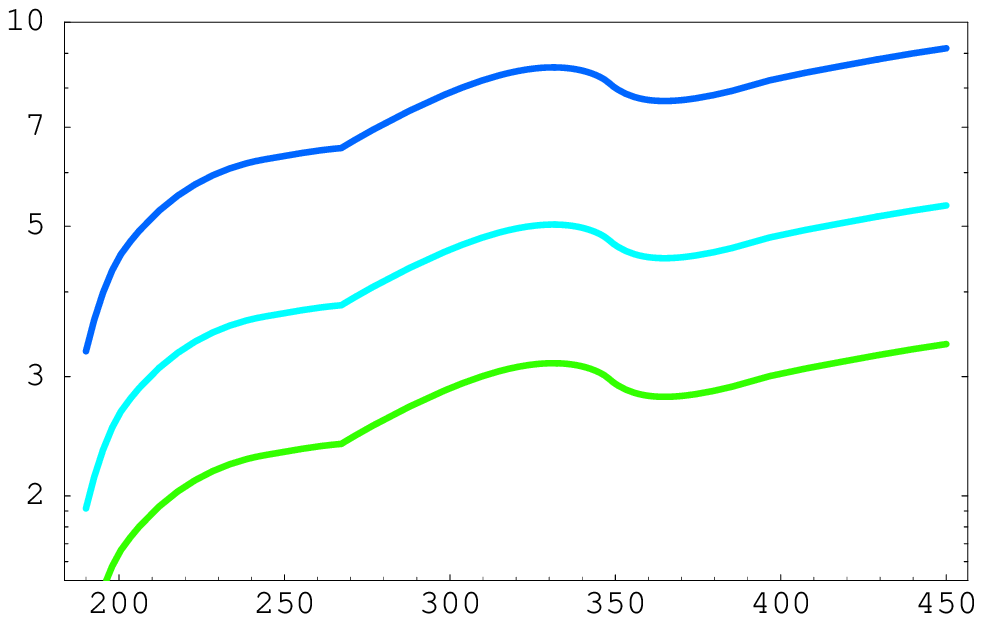}}
\caption{Signal significance of $ZZ$ channel as a 
function of sgoldstino mass $m_S$ for the model~II.
\label{fig:z-low-2}}
\begin{picture}(0,0)(-38,72)
\put(-60.00,350.00){\makebox(0,0)[cb]{\LARGE$\frac{N_{\rm
S}}{\sqrt{N_{\rm B}}}$}}
\put(40.00,355.00){\makebox(0,0)[cb]{${\cal L}=100$~fb$^{-1}$}}
\put(320.00,335.00){\makebox(0,0)[cb]{\large$\sqrt{F}=1.75$~TeV}}
\put(320.00,270.00){\makebox(0,0)[cb]{\large$\sqrt{F}=2$~TeV}}
\put(320.00,210.00){\makebox(0,0)[cb]{\large$\sqrt{F}=2.25$~TeV}}
\put(200.00,115.00){\makebox(0,0)[cb]{\large$m_S$,~GeV}}
\end{picture}
\end{figure}

\begin{figure}
\centerline{\epsfxsize=1.0\textwidth \epsfbox{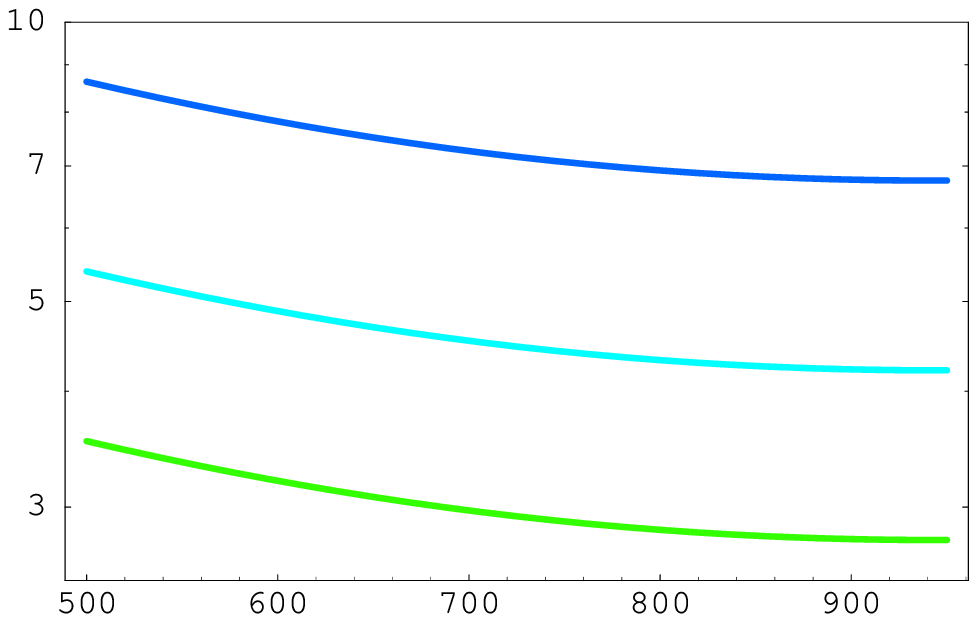}}
\caption{Signal significance of $ZZ$ channel 
as a function of sgoldstino mass $m_S$ for the model~II.
\label{fig:z-high-2}}
\begin{picture}(0,0)(-38,-246)
\put(50.00,-10.00){\makebox(0,0)[cb]{${\cal L}=100$~fb$^{-1}$}}
\put(-50.00,35.00){\makebox(0,0)[cb]{\LARGE$\frac{N_{\rm
S}}{\sqrt{N_{\rm B}}}$}}
\put(320.00,7.00){\makebox(0,0)[cb]{\large$\sqrt{F}=2$~TeV}}
\put(320.00,-75.00){\makebox(0,0)[cb]{\large$\sqrt{F}=2.25$~TeV}}
\put(320.00,-150.00){\makebox(0,0)[cb]{\large$\sqrt{F}=2.5$~TeV}}
\put(200.00,-202.00){\makebox(0,0)[cb]{\large$m_S$,~GeV}}
\end{picture}
\end{figure}

Our main results are the  plots 
7 - 12 
with the LHC sensitivity to the scale of
supersymmetry breaking estimated by making use of 
Eq.~(\ref{estimate-of-significance}):
Figures~\ref{fig:photon-1},\ref{fig:photon-2} refer to two-photon channel,  
Figures~\ref{fig:z-low-1},\ref{fig:z-low-2} 
correspond to $ZZ$ channel and $M_S<500$~GeV, 
Figures~\ref{fig:z-high-1},\ref{fig:z-high-2} 
concern $ZZ$ channel and $M_S>500$~GeV. 


We would like to mention
that we did not take into account 
sgoldstino couplings to superpartners. 
If open (allowed kinematically), new 
sgoldstino decay channels distort the pattern of sgoldstino branching 
ratios into SM particles. In a given model the level of distortion
depends on the set of soft supersymmetry breaking masses and
sgoldstino masses and effective couplings to superpartners. 
As clearly seen from 
Eq.~(\ref{estimate-of-significance}), the distortion can affect 
our predictions of LHC sensitivity iff a partial width of some new channel
becomes comparable to the partial width of the dominant gluonic
channel. The deviation of the
predictions may be estimated from
Eq.~(\ref{estimate-of-significance}). In particular, if the total
decay rate into superpartners becomes equal to decay rate into two
gluons, the sensitivity to $\sqrt{F}$ decreases by 20\%.  

It should be noted that in  Ref.~\cite{Giudice:2000av} 
by making use of the same method as our the sensitivity 
of the LHC to the radion has been investigated. 
Certainly the method is viable for any scalar SM singlet
massive field.                             
 
The work is supported in part by 
CPG and SSLSS grant 00-15-96626.  The work of D.G.\ is also supported
in part by the RFBR grants 01-02-16710, 02-02-17398, 
by INTAS YSF 2001/2-142 and by
the program SCOPES of 
the Swiss National Science Foundation, project No.~7SUPJ062239. 
The work of N.K.\ is also supported by CERN-INTAS 99-0377.

\newpage


\begin{thebibliography}{99}

\bibitem{ellis}
J.~R.~Ellis, K.~Enqvist and D.~V.~Nanopoulos,
Phys.\ Lett.\ B {\bf 147} (1984) 99;
J.~R.~Ellis, K.~Enqvist and D.~V.~Nanopoulos,
Phys.\ Lett.\ B {\bf 151} (1985) 357.

\bibitem{gmm}
G.~F.~Giudice and R.~Rattazzi,
Phys.\ Rept.\  {\bf 322} (1999) 419
[arXiv:hep-ph/9801271];
S.~L.~Dubovsky, D.~S.~Gorbunov and S.~V.~Troitsky,
Phys.\ Usp.\  {\bf 42} (1999) 623
[Usp.\ Fiz.\ Nauk {\bf 169} (1999) 705]
[arXiv:hep-ph/9905466].

\bibitem{Gorbunov:2001pd}
D.~S.~Gorbunov and A.~V.~Semenov,
LAPTH-884/01, 
[arXiv:hep-ph/0111291].

\bibitem{Perazzi:2000id}
E.~Perazzi, G.~Ridolfi and F.~Zwirner,
Nucl.\ Phys.\ B {\bf 574} (2000) 3
[arXiv:hep-ph/0001025].

\bibitem{Perazzi:2000ty} 
E.~Perazzi, G.~Ridolfi and F.~Zwirner,
Nucl.\ Phys.\ B {\bf 590} (2000) 287
[arXiv:hep-ph/0005076].

\bibitem{Nowakowski:1994ag} 
M.~Nowakowski and S.~D.~Rindani,
Phys.\ Lett.\ B {\bf 348} (1995) 115
[arXiv:hep-ph/9410262].

\bibitem{Gorbunov:2000th} 
D.~S.~Gorbunov,
Nucl.\ Phys.\ B {\bf 602} (2001) 213
[arXiv:hep-ph/0007325].

\bibitem{Dicus:1989gg} 
D.~A.~Dicus, S.~Nandi and J.~Woodside,
Phys.\ Rev.\ D {\bf 41} (1990) 2347.

\bibitem{Dicus:1990su} 
D.~A.~Dicus and P.~Roy,
Phys.\ Rev.\ D {\bf 42} (1990) 938.

\bibitem{Dicus:1990dy} 
D.~A.~Dicus, S.~Nandi and J.~Woodside,
Phys.\ Rev.\ D {\bf 43} (1991) 2951.

\bibitem{Dicus:1996ua}
D.~A.~Dicus and S.~Nandi,
Phys.\ Rev.\ D {\bf 56} (1997) 4166
[arXiv:hep-ph/9611312].

\bibitem{Abreu:2000ij} 
P.~Abreu {\it et al.}  [DELPHI Collaboration],
Phys.\ Lett.\ B {\bf 494} (2000) 203
[arXiv:hep-ex/0102044].

\bibitem{CMS-TP} ``CMS:  
Technical propose'', CERN/LHCC/94-38, 15 december 1994.

\bibitem{signif-higgs-gamma-gamma} 
``CMS: The electromagnetic calorimeter. Technical design report'',  
CERN-LHCC-97-33, 15 december 1997.


\bibitem{Denegri:1995fr}
D.~Denegri  [CMS Collaboration],
CERN-PPE-95-183
{\it Invited talk at the Conf. on Elementary Particle Physics, Present
and Future, Valencia, Spain, Jun 5-8, 1995}.

\bibitem{Giudice:2000av}
G.~F.~Giudice, R.~Rattazzi and J.~D.~Wells,
Nucl.\ Phys.\ B {\bf 595} (2001) 250
[arXiv:hep-ph/0002178].
\end{thebibliography}
\end{document}